\documentclass[10pt]{iopart}

\usepackage{iopams}
\usepackage{graphicx} 
\usepackage{dcolumn}
\usepackage{times}
\usepackage{textcomp}
\usepackage{titlesec}
\usepackage{float}
\usepackage{color,xcolor}
\usepackage{graphicx,times} 
\usepackage{cuted} \stripsep -3pt plus 3pt minus 2pt
\usepackage{epstopdf}
\usepackage{graphicx}
\usepackage{subfigure}
\usepackage{makecell}
\usepackage{multirow}
\renewcommand{\figurename}{Figure}
\usepackage[colorlinks,linkcolor=blue, urlcolor=blue, anchorcolor=blue, citecolor=blue]{hyperref}
\usepackage{longtable,booktabs}
\usepackage{fouriernc}
\usepackage{epsfig,graphics,graphicx}
\usepackage{bm}
\usepackage{color}
\usepackage{flushend}
\usepackage{ulem}
\usepackage[sort&compress,square,numbers]{natbib}
\usepackage[final]{pdfpages}

\begin{document}
\title{On the Upper Bound of Non-Thermal Fusion Reactivity with Fixed Total Energy}

\author{
Huasheng Xie$^{1,2}$
and Xueyun Wang$^{1,2}$
}

\address{

$^1$ Hebei Key Laboratory of Compact Fusion, Langfang 065001, People’s Republic of China

$^2$ ENN Science and Technology Development Co., Ltd., Langfang 065001, People’s Republic of China

}
\eads{\mailto{huashengxie@gmail.com, xiehuasheng@enn.cn}}

\begin{indented}
\item[\today]
\end{indented}

\begin{abstract}
Fusion reactivity represents the integration of fusion cross-sections and the velocity distributions of two reactants. In this study, we investigate the upper bound of fusion reactivity for a non-thermal reactant coexisting with a thermal Maxwellian background reactant while maintaining a constant total energy. Our optimization approach involves fine-tuning the velocity distribution of the non-thermal reactant. We employ both Lagrange multiplier and Monte Carlo methods to analyze Deuterium-Tritium (D-T) and Proton-Boron11 (p-B11) fusion scenarios. Our findings demonstrate that, within the relevant range of fusion energy, the maximum fusion reactivity can often surpass that of the conventional Maxwellian-Maxwellian reactants case by a substantial margin, ranging from 50\% to 300\%. These enhancements are accompanied by distinctive distribution functions for the non-thermal reactant, characterized by one or multiple beams. These results not only establish an upper limit for fusion reactivity but also provide valuable insights into augmenting fusion reactivity through non-thermal fusion, which holds particular significance in the realm of fusion energy research.
\end{abstract}
\maketitle
\ioptwocol

\section{Introduction}\label{sec:intro}

The counting of fusion reactions per unit volume and per unit time is given by\cite{Atzeni2004,Clayton1983}:
\begin{equation}
R_{12}=\frac{n_1n_2}{1+\delta_{12}}\langle\sigma v\rangle,
\end{equation}
where $n_1$ and $n_2$ represent the number densities of the two reactants, respectively. The term $\delta_{12}$ is equal to 0 for different reactants and equal to 1 for the same reactants, in order to prevent double counting of the reaction.

Here, $\sigma=\sigma(E)$ or $\sigma=\sigma(v)$ represents the fusion cross section, with $E$ being the energy in the center-of-mass frame, defined as
\begin{equation}
E=\frac{1}{2}m_rv^2,~~v=|{\bm v}|=|{\bm v}_1-{\bm v}_2|,~~m_r=\frac{m_1m_2}{m_1+m_2},
\end{equation}
where $m_1$ and $m_2$ denote the mass of the two reactants, and $m_r$ represents the reduced mass of the system. The fusion reactivity $\langle\sigma v\rangle$ is calculated as the integral of the fusion cross section and the velocity distribution functions of the reactants:
\begin{equation}\label{eq:sgmv}
\langle\sigma v\rangle=\int\int d{\bm v}_1d{\bm v}_2\sigma(|{\bm v}_1-{\bm v}_2|)|{\bm v}_1-{\bm v}_2|f_1({\bm v}_1)f_2({\bm v}_{ 2}),
\end{equation}
where $f_1$ and $f_2$ are the normalized velocity distribution functions of the two ions, i.e., $\int f_{j}({\bm v}_j)d{\bm v}_j=1$ with $j=1,2$, and $d{\bm v}_j=dv_{xj}dv_{yj}dv_{zj}$. We assume $m_1\leq m_2$, i.e., the mass of the first reactant is lighter than the second one.

It is interesting to study non-thermal fusion reactivity, as it is common in fusion experiments (cf., \cite{Hartouni2023}), and it offers a potentially attractive solution to enhance realistic fusion energy production with the challenges posed by advanced fuels (cf., \cite{Nevins1998, Cai2022, Ochs2022, Mehlhorn2022, Rostoker1997}). Comprehensive theoretical investigations of fusion reactivities involving common drift bi-Maxwellian distributions \cite{Xie2023}, drift ring beams, slowing down, and superthermal kappa distributions \cite{Kong2023}, and velocity-space
anisotropy distribution\cite{Kolmes2021} in fusion plasmas have revealed potential enhancement ranges for fusion reactivity when compared to thermal Maxwellian plasmas. It has also been demonstrated that even a modest increase in proton-Boron fusion reactivity, such as 20\%, can significantly impact the feasibility of proton-Boron fusion energy production (cf., \cite{Xie2023, Putvinski2019}). { The growing interest in studying advanced fuel for fusion energy \cite{Dawson1981,Liu2023} requires the exploration of potential approaches to increase fusion reactivity and reduce radiation loss. Therefore, enhancing fusion reactivity becomes a crucial topic that needs resolution.}

In our previous studies (cf., \cite{Xie2023, Kong2023}), the fusion reactivity enhancement factors (defined in Eq. (\ref{eq:fsgmv})) for several typical non-thermal distributions mainly fell within the range of 0.5-1.5. While we were able to calculate the fusion reactivity for arbitrary ion velocity distributions using a simple and fast approach (cf., \cite{Xie2023a}), it did not provide information about which distribution could yield the maximum fusion reactivity or what the upper limit of the enhancement factor might be. This question has motivated the present work. Resolving this problem can provide us with insights into selecting specific distributions to maximize fusion yields.

{In fusion energy research, the most crucial parameter is the ratio of output fusion yield energy to input heating energy. The input heating energy is associated with the total energy of the plasma.} The total energy for a given distribution is $E_j=\frac{1}{2}m_j\int v_j^2f_j({\bm v}_j)d{\bm v}_j$. In this work, we are interested in determining the maximum fusion reactivity $\langle\sigma v\rangle$ when $E_j$ is fixed, along with the corresponding distribution functions $f_j$. To make the problem more analytically tractable, we limit $f_2$ to be a Maxwellian (thermal) distribution and focus on optimizing only $f_1({\bm v})$.

In Section \ref{sec:method}, we present the strategy we employ to address this issue. In Section \ref{sec:result}, we apply our strategy to study D-T (Deuterium-Tritium) and p-$^{11}{\rm B}$ (proton-Boron) fusion and present the results. Finally, in Section \ref{sec:summ}, we summarize our findings.

\section{Strategy to Simplify the Problem and Numerical Investigations}\label{sec:method}

Our objective is to determine the maximum value of $\langle\sigma v\rangle$ for a given energy of the first reactant $E_1$ and the temperature of the second reactant $T_2$.

Optimizing the three-dimensional distribution $f_1({\bm v})=f_1(v,\theta,\phi)$ can be challenging. However, if we express $f_1({\bm v})$ as a series of delta functions, as follows:
\begin{equation}\label{eq:f13d}
f_1({\bm v})=\frac{1}{N}\sum_{n=1}^{N}\delta({\bm v}-{\bm v}_n),
\end{equation}
the problem becomes more tractable. As $N\to\infty$, this form of $f_1$ can be used to approximate arbitrary distribution functions.

\begin{figure*}
\centering
\includegraphics[width=14cm]{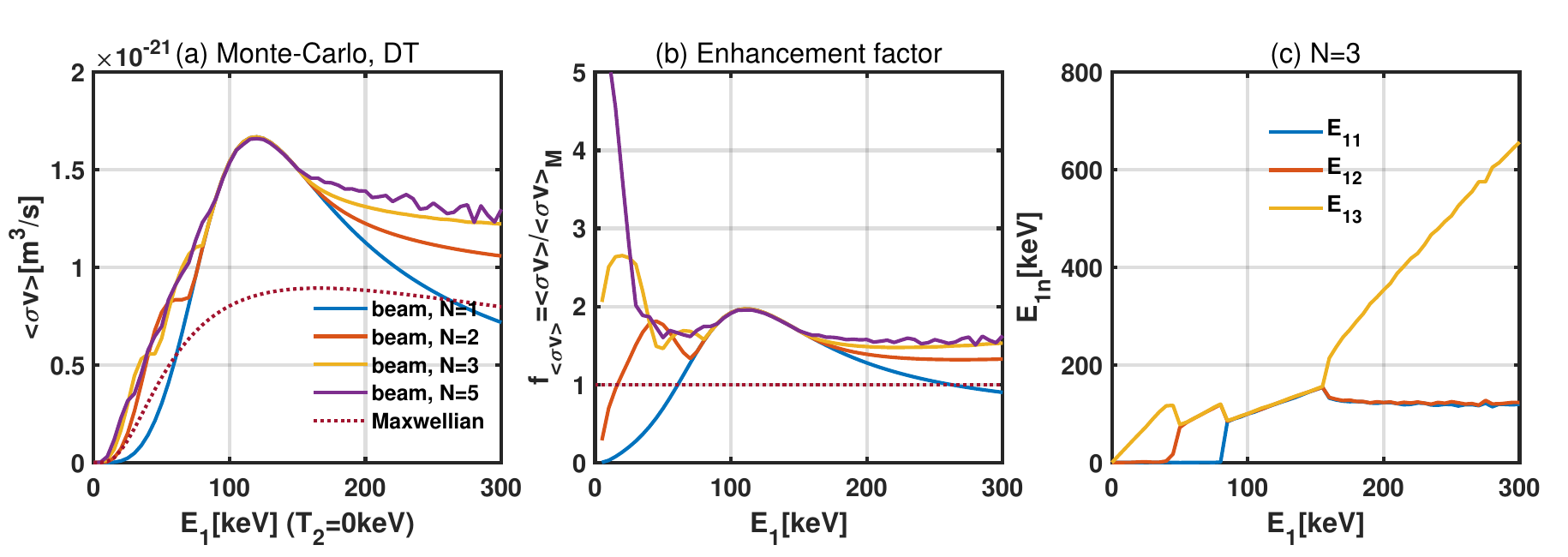}
\caption{Monte-Carlo numerical investigation of the maximum fusion reactivity of D-T with temperature $T_2=0$ for different $N$ beams. { (a) Reactivity for different $N$ and for Maxwellian cases. (b) Enhancement factors for different $N$ compared to the Maxwellian case. (c) For the $N=3$ case, Monte-Carlo energies of each beams when the reactivity is at its maximum.}}\label{fig:maxsgmvmc_DT}
\end{figure*}

\begin{figure*}
\centering
\includegraphics[width=14cm]{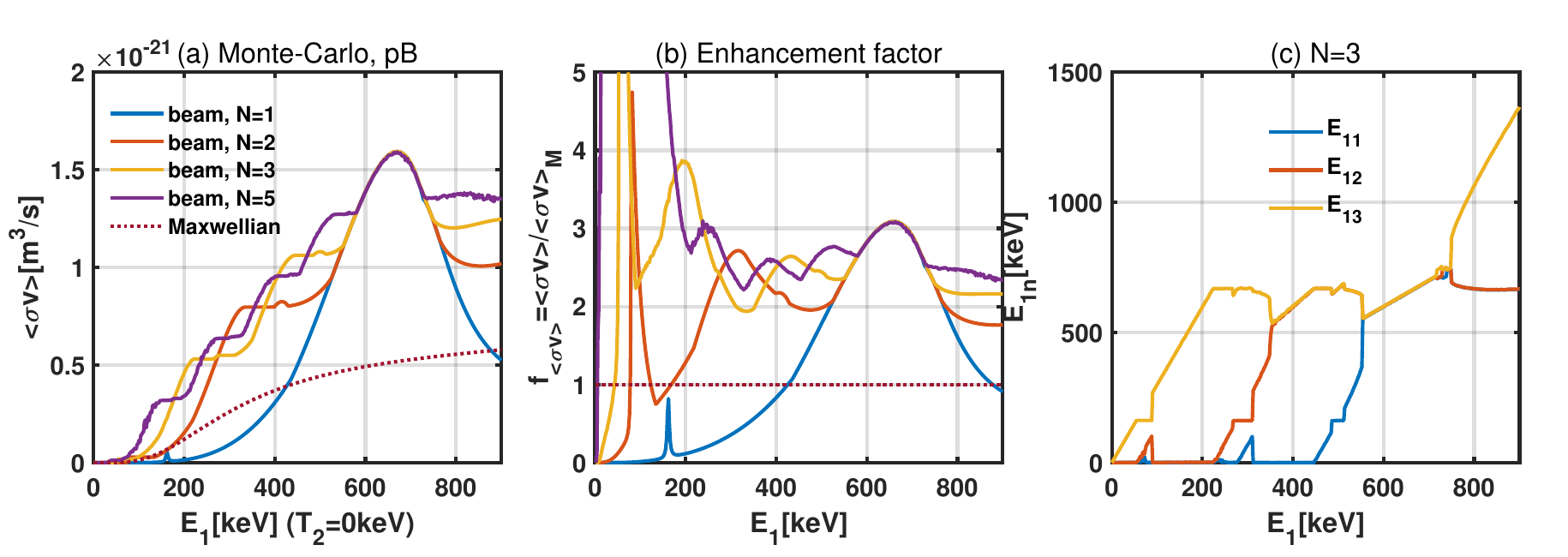}
\caption{Monte-Carlo numerical investigation of the maximum fusion reactivity of p-B with temperature $T_2=0$ for different $N$ beams. { (a) Reactivity for different $N$ and for Maxwellian cases. (b) Enhancement factors for different $N$ compared to the Maxwellian case. (c) For the $N=3$ case, Monte-Carlo energies of each beams when the reactivity is at its maximum.}}\label{fig:maxsgmvmc_pB}
\end{figure*}

\begin{figure*}
\centering
\includegraphics[width=14cm]{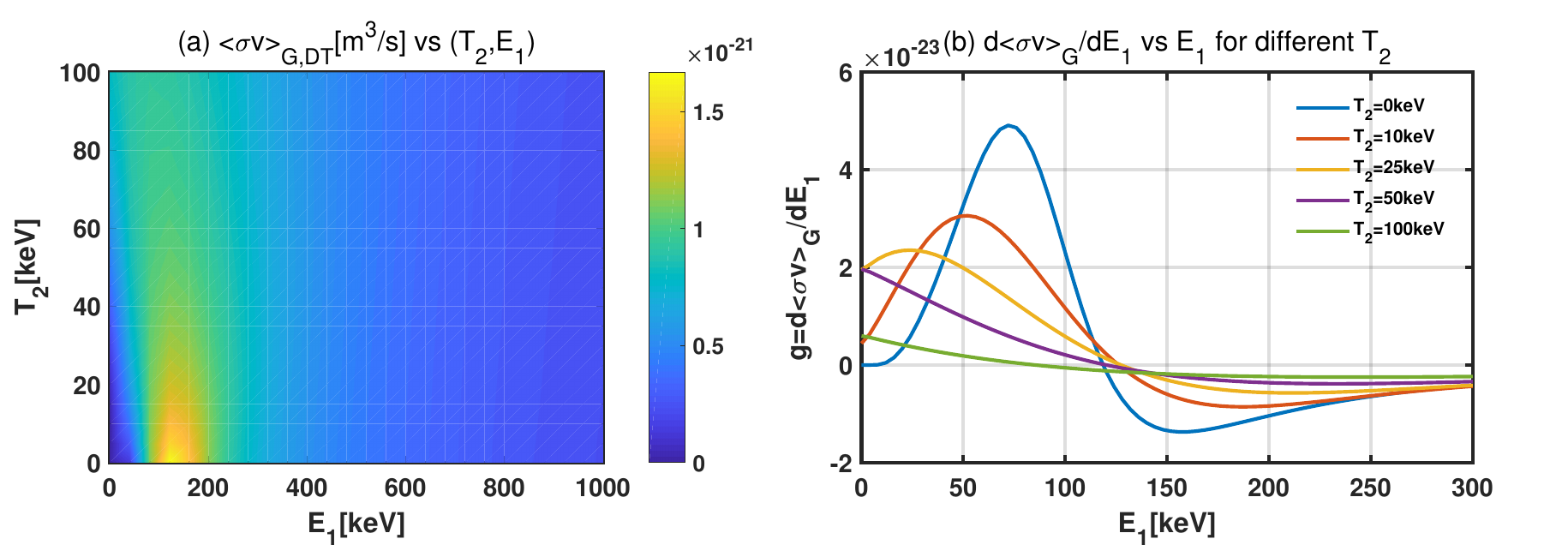}
\caption{Fusion reactivity ${\langle\sigma v\rangle}_G$ and $g=\frac{d{\langle\sigma v\rangle}_G}{dE_{1n}}$ of beam-Maxwellian distributions for D-T vs temperature $T_2$ and drift energy $E_{1}$.}\label{fig:sgmvj_DT}
\end{figure*}

\begin{figure*}
\centering
\includegraphics[width=14cm]{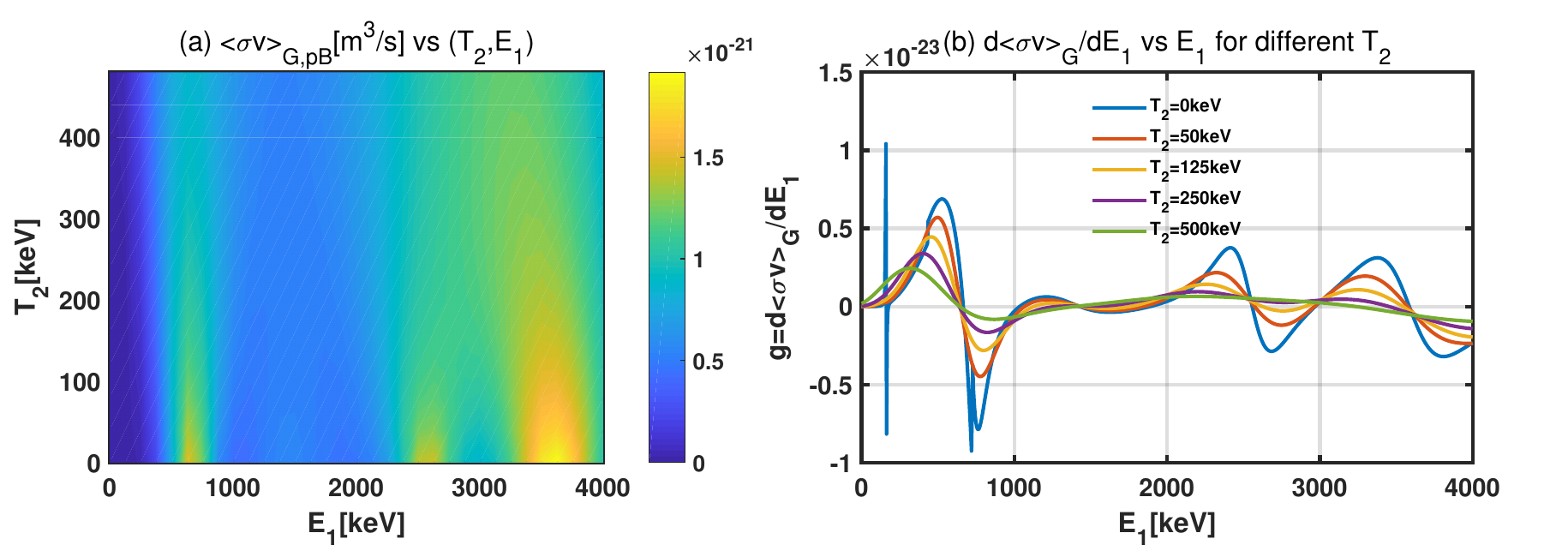}
\caption{Fusion reactivity ${\langle\sigma v\rangle}_G$ and $g=\frac{d{\langle\sigma v\rangle}_G}{dE_{1n}}$ of beam-Maxwellian distributions for p-B vs temperature $T_2$ and drift energy $E_{1}$.}\label{fig:sgmvj_pB}
\end{figure*}

\subsection{Reducing the Dimensions of the Problem}
Initially, considering that $f_2$ is an isotropic Maxwellian distribution background given by
\begin{equation}\label{eq:fm}
f_2({\bm v})=\Big(\frac{m_2}{2\pi k_BT_2}\Big)^{3/2}\exp\Big[-\frac{m_2{\bm v}^2}{2k_BT_2}\Big],
\end{equation}
we observe that each beam injected into the Maxwellian background results in fusion reactivity that is solely dependent on the beam's energy, independent of its direction. This can also be inferred from Eq.(\ref{eq:sgmvg}) in \ref{sec:sgmvg}. Thus, for the study of fusion reactivity, we can limit our consideration to the one-dimensional distribution of $f_1$, specifically:
\begin{equation}\label{eq:f11d}
f_1(v)=\frac{1}{N}\sum_{n=1}^{N}\delta(v-v_n),
\end{equation}
where $f_1(v)=\int_0^{\pi} d\theta \int_0^{2\pi} d\phi v^2\sin\theta f_1({\bm v})$, and $v=|{\bm v}|$, $v_n=|{\bm v}_n|\geq0$. { Notice that the fusion reactivity of two drift Maxwellian reactants forms the basis for the present work, we provide the details in \ref{sec:sgmvg}.}

Hence, we have proven that $f_1$ can be expressed as $f_1(v)=\frac{1}{N}\sum_{n=1}^{N}\delta(v-v_n)$. The problem then becomes the optimization of $v_n$ to maximize:
\begin{eqnarray}\label{eq:sumsgmvg0}
\langle\sigma v\rangle=\frac{1}{N}\sum_{n=1}^{N}\langle\sigma v\rangle_{Gn},
\end{eqnarray}
where $\langle\sigma v\rangle_{Gn}$ can be calculated from Eq.(\ref{eq:sgmvg}). The constraint is:
\begin{eqnarray}\label{eq:constrain0}
\frac{1}{2}m_1\frac{1}{N}\sum_{n=1}^{N}v_n^2=E_{1}.
\end{eqnarray}

Solving the above problem in energy space is much simpler than in velocity space. We set $f_1(E)=\frac{1}{N}\sum_{n=1}^{N}\delta(E-E_{1n})$, and the problem becomes the optimization of $E_{1n}=\frac{1}{2}m_1v_n^2$ to maximize:
\begin{eqnarray}\label{eq:sumsgmvg}
\langle\sigma v\rangle=\frac{1}{N}\sum_{n=1}^{N}\langle\sigma v\rangle_{Gn},
\end{eqnarray}
with the constraint:
\begin{eqnarray}\label{eq:constrain}
\frac{1}{N}\sum_{n=1}^{N}E_{1n}=E_{1}.
\end{eqnarray}

The fusion reactivity enhancement factor is defined as:
\begin{eqnarray}\label{eq:fsgmv}
f_{\langle\sigma v\rangle}=\frac{\langle\sigma v\rangle}{\langle\sigma v\rangle_M},
\end{eqnarray}
where $\langle\sigma v\rangle_M$ is the fusion reactivity of the corresponding Maxwellian case with the same kinetic energy. Note that for the drift Maxwellian distribution in Eq.(\ref{eq:fbm}), the total kinetic energy of species 1 is\cite{Xie2023}: $E_{k1}=\frac{1}{2}m_1\int v_1^2f_1({\bm v}_1)d{\bm v}_1=\frac{3}{2}k_BT_1+E_{d1}$. Hence, to maintain the total kinetic energy the same, the corresponding Maxwellian temperature for the beam energy is $k_BT_1=\frac{2}{3}E_1$.

\begin{figure*}
\centering
\includegraphics[width=13cm]{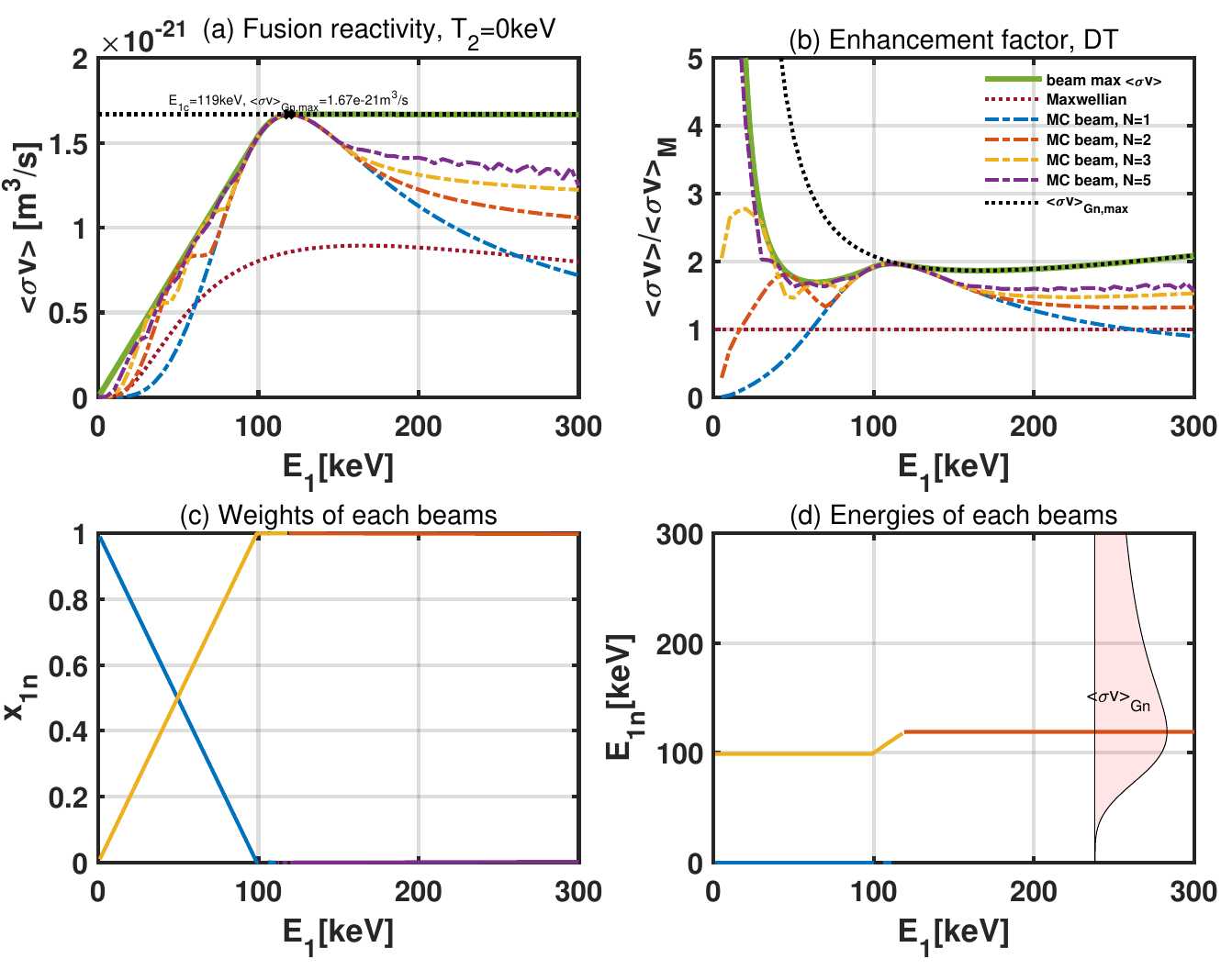}
\caption{Upper limit of D-T fusion reactivity ${\langle\sigma v\rangle}$ vs drift energy $E_{1}$, with temperature $T_2=0$, { using the method outlined in subsection \ref{sec:Lag} numerically}.  { (a) Fusion reactivity. (b) Enhancement factor. (c) Weights $x_{1n}$ of each beams for the upper limit case. (d) Energies $E_{1n}$ of each beams for the upper limit case. The $\langle\sigma v\rangle_{Gn}$ vs $E_{1n}$ data is also shown.}}\label{fig:solve_maxsgmvlinprog_icase_1_T2_0}
\end{figure*}

\begin{figure*}
\centering
\includegraphics[width=13cm]{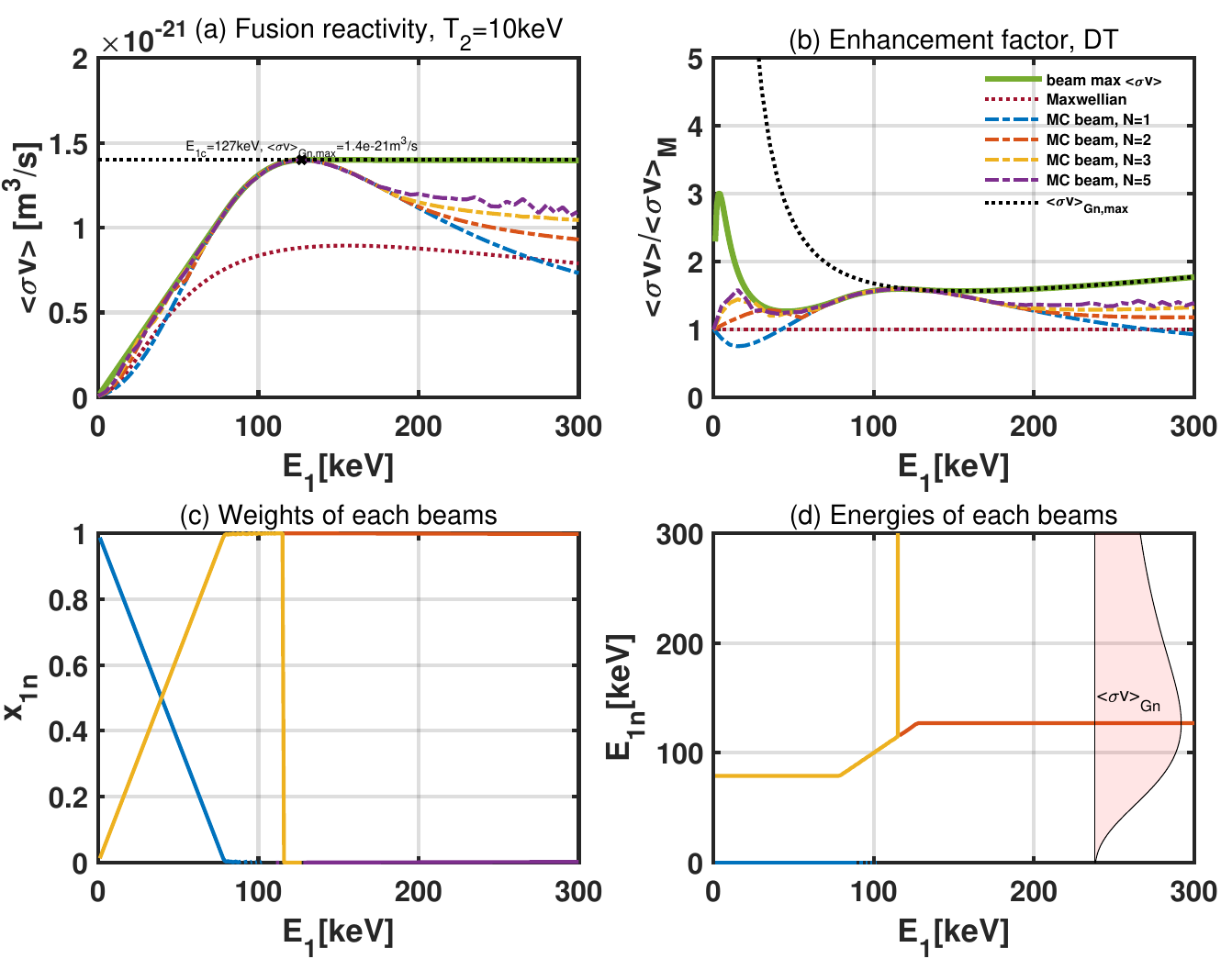}
\caption{Upper limit of D-T fusion reactivity ${\langle\sigma v\rangle}$ vs drift energy $E_{1}$, with temperature $T_2=10$keV, { using the method outlined in subsection \ref{sec:Lag} numerically}. { (a) Fusion reactivity. (b) Enhancement factor. (c) Weights $x_{1n}$ of each beams for the upper limit case. (d) Energies $E_{1n}$ of each beams for the upper limit case. The $\langle\sigma v\rangle_{Gn}$ vs $E_{1n}$ data is also shown.}}\label{fig:solve_maxsgmvlinprog_icase_1_T2_10}
\end{figure*}

\begin{figure*}
\centering
\includegraphics[width=13cm]{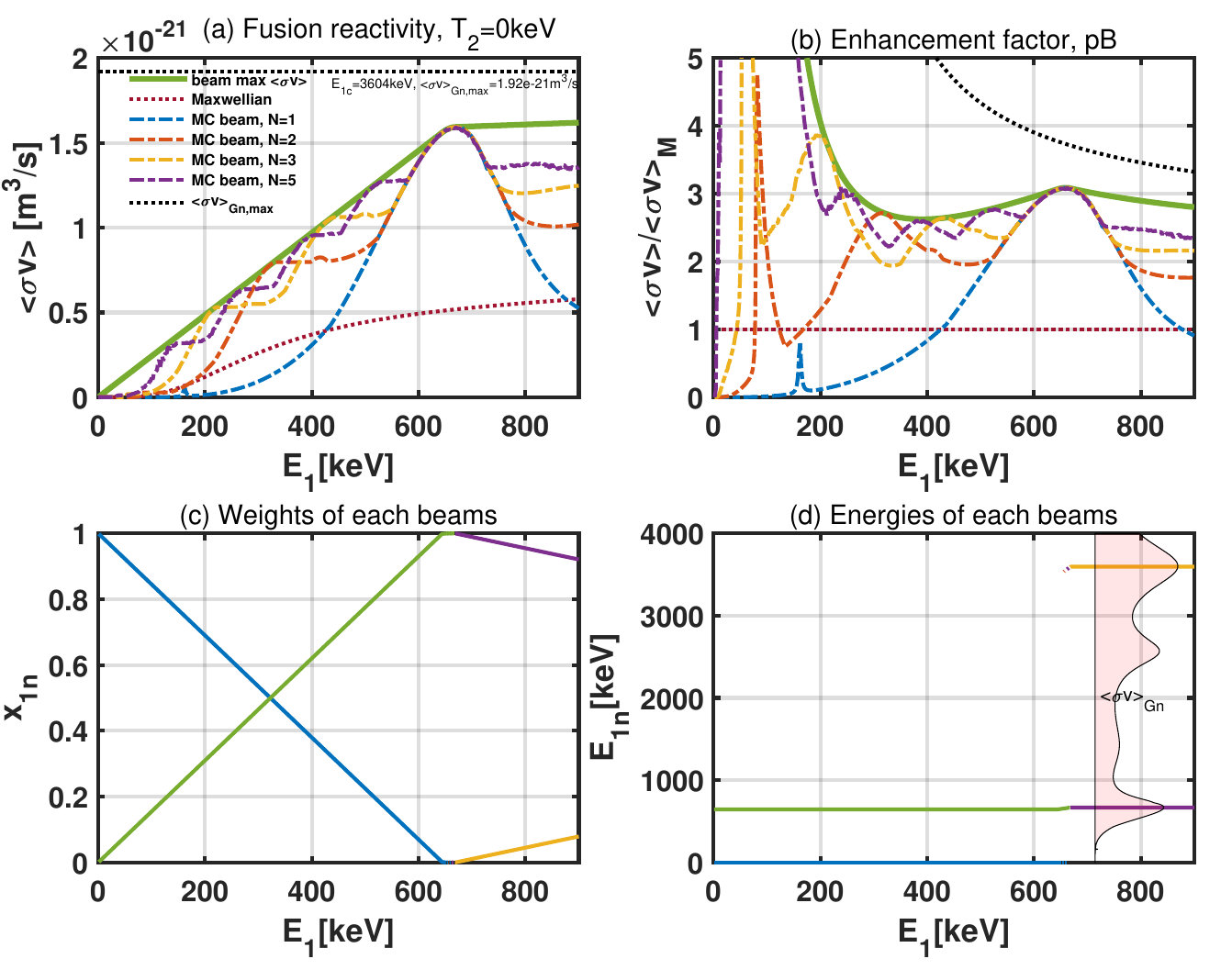}
\caption{Upper limit of p-B fusion reactivity ${\langle\sigma v\rangle}$ vs drift energy $E_{1}$, with temperature $T_2=0$, { using the method outlined in subsection \ref{sec:Lag} numerically}. { (a) Fusion reactivity. (b) Enhancement factor. (c) Weights $x_{1n}$ of each beams for the upper limit case. (d) Energies $E_{1n}$ of each beams for the upper limit case. The $\langle\sigma v\rangle_{Gn}$ vs $E_{1n}$ data is also shown.}}\label{fig:solve_maxsgmvlinprog_icase_4_T2_0}
\end{figure*}

\begin{figure*}
\centering
\includegraphics[width=13cm]{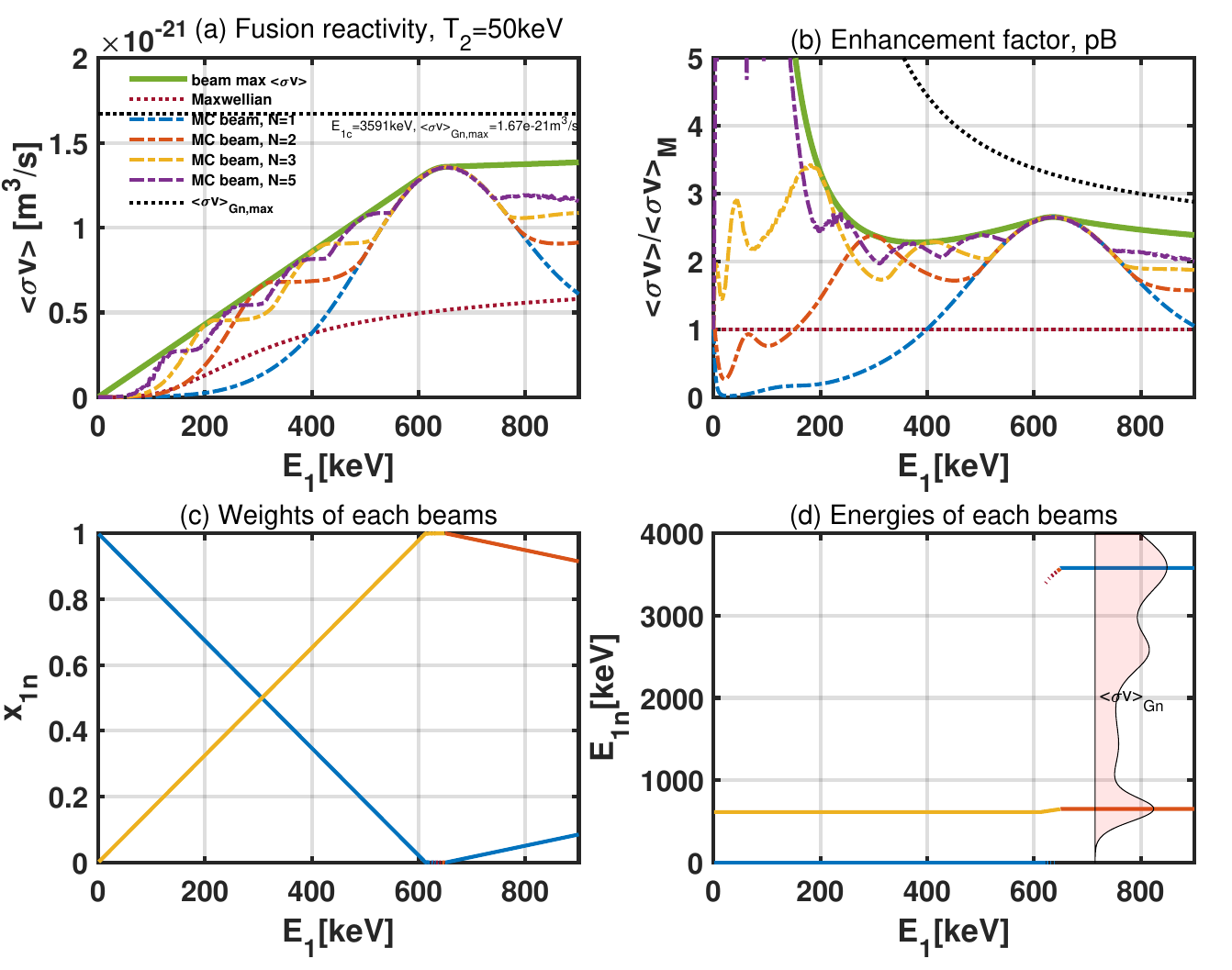}
\caption{Upper limit of p-B fusion reactivity ${\langle\sigma v\rangle}$ vs drift energy $E_{1}$, with temperature $T_2=50$keV, { using the method outlined in subsection \ref{sec:Lag} numerically}. { (a) Fusion reactivity. (b) Enhancement factor. (c) Weights $x_{1n}$ of each beams for the upper limit case. (d) Energies $E_{1n}$ of each beams for the upper limit case. The $\langle\sigma v\rangle_{Gn}$ vs $E_{1n}$ data is also shown.}}\label{fig:solve_maxsgmvlinprog_icase_4_T2_50}
\end{figure*}

\subsection{Monte-Carlo numerical investigation}\label{sec:mc}
Before studying the general cases, we investigate cases with small values of $N$ and $T_2=0$. We randomly choose $E_{1n}$ to find the maximum of Eq.(\ref{eq:sumsgmvg}), while satisfying Eq.(\ref{eq:constrain}). In Figures \ref{fig:maxsgmvmc_DT} and \ref{fig:maxsgmvmc_pB}, it is evident that for $N=1$, the fusion reactivity is not always greater than that of the Maxwellian case. However, for $N=2$, it can be higher. When $N=5$, it approaches the $N=3$ case at $T_1>25$ keV for D-T and $T_1>200$ keV for p-B. The fusion cross-section data are obtained from \cite{Bosch1992} for D-T and \cite{Tentori2023,Sikora2016} for p-B.

Figure \ref{fig:maxsgmvmc_DT} also shows that the maximum $f_{\langle\sigma v\rangle}\simeq1.5-2$ for D-T fusion in the energy range $E_1\in[40,300]$ keV. Figure \ref{fig:maxsgmvmc_pB} shows that the maximum $f_{\langle\sigma v\rangle}\simeq2-3$ for p-B fusion in the energy range $E_1\in[200,900]$ keV. { For the lower energy range of $E_1$, the maximum $f_{\langle\sigma v\rangle}$ can be higher, which is understandable since the benefit from the peak of the cross section can be more significant. In the typical D-T fusion energy research range, $E_1={\frac{3}{2}}k_BT_1\simeq15-30$keV\cite{Xie2023}, the maximum enhancement factor can be $f_{\langle\sigma v\rangle}\simeq5$.} To achieve the maximum $f_{\langle\sigma v\rangle}$, the distribution function of $f_1(E)$ typically consists of one or two energetic beams with a cold background, for example, $E_{11}=E_{12}=0$ and $E_{13}\neq0$ for $N=3$. The similarity between the $N=3$ and $N=5$ cases suggests that having $3-5$ beams is sufficient to approach the upper limit. Due to the weight of each beam being $\frac{1}{N}$, the maximum reactivity for larger values of $N$ is not always greater than that for smaller values of $N$.

This Monte-Carlo investigation provides us with a rough but intuitive understanding of how large $f_{\langle\sigma v\rangle}$ can be and the possible beam energies. However, it is challenging to determine the rigorous theoretical upper limit of $f_{\langle\sigma v\rangle}$ since we cannot calculate for $N\to\infty$, especially to obtain accurate solutions, { which is the topic of subsection \ref{sec:Lag}}.

\subsection{Theoretical maximum via Lagrange multipler method}\label{sec:Lag}

We can utilize the Lagrange multiplier method to determine the maximum of Eq.(\ref{eq:sumsgmvg}) while adhering to the constraint given in Eq.(\ref{eq:constrain}). We define the Lagrangian as follows:
\begin{eqnarray}\label{eq:Lag}\nonumber
&&L(E_{11},E_{12},\cdots,E_{1N},\lambda)\\
&=&\frac{1}{N}\sum_{n=1}^{N}\langle\sigma v\rangle_{Gn}-\lambda\Big[\frac{1}{N}\sum_{n=1}^{N}E_{1n}-E_{1}\Big],
\end{eqnarray}
which leads to the requirement:
\begin{eqnarray}\label{eq:dLag}
\frac{\partial L(E_{11},E_{12},\cdots,E_{1N},\lambda)}{\partial E_{1n}}=\frac{1}{N}\frac{\partial \langle\sigma v\rangle_{Gn}}{\partial E_{1n}}-\lambda \frac{1}{N}=0.
\end{eqnarray}
Eq.(\ref{eq:dLag}) has solutions where either $E_{1n}=0$ or 
\begin{eqnarray}\label{eq:g}
g(E_{1n},T_2)\equiv \frac{\partial \langle\sigma v\rangle_{Gn}}{\partial E_{1n}}=\lambda,
\end{eqnarray}
holds for all $n=1,2,\cdots, N$, with $\lambda$ being a constant. For a given $T_2$, we observe in Figures \ref{fig:sgmvj_DT} and \ref{fig:sgmvj_pB} that the function $g(E_{1n},T_2)={constant}$ can only be satisfied for a few specific $E_{1n}$ values. For example, in the case of D-T fusion, there are only 1-2 solutions when $T_2=0$. This implies that the maximum of Eq.(\ref{eq:sumsgmvg}) occurs when $E_{1n}$ takes on 1-3 specific values, indicating that the distribution of $f_1(E)$ should consist of 1-3 beams, regardless of the value of $N$. The actual values of the beam energies depend on Eq.(\ref{eq:constrain}).

The model described above remains challenging to solve for $N\to\infty$. Nevertheless, armed with the insights gained from the previous results, we can make further progress. 
{ Since only a limited number of solutions exist from the Lagrange multiplier approach, i.e., even as $N\to\infty$, the energies of these solutions are confined to several specific values, such as $N_1$ for energy $E_1$, $N_2$ for energy $E_2$, and so on, where $N=N_1+N_2+N_3$ (take 3 solutions as an example). Therefore, it is unnecessary to solve for  $N_1$, $N_2$ and $N_3$ individually as $N\to\infty$. Instead, we further assume $x_1=N_1/N$, $x_2=N_2/N$, $x_3=N_3/N$, and solve for $x_1$, $x_2$, $x_3$, which can yield rigorous solutions.}

{  Now, we present the new procedure.} Let us assume
\begin{equation}
f_1(E)=\sum_{n=1}^{N}x_n\delta(E-E_{1n}),
\end{equation}
where $N$ represents the number of solutions to Eq.(\ref{eq:dLag}), and $x_n$ denotes the weight of the corresponding beam. This assumption is subject to the constraints:
\begin{equation}\label{eq:solveLag1}
\sum_{n=1}^{N}x_n=1,~~0\leq x_n\leq1,
\end{equation}
and
\begin{equation}\label{eq:solveLag2}
\sum_{n=1}^{N}x_nE_{1n}=E_1.
\end{equation}
We aim to maximize
\begin{eqnarray}\label{eq:solveLag3}
\langle\sigma v\rangle=\sum_{n=1}^{N}x_n\langle\sigma v\rangle_{Gn}.
\end{eqnarray}

Given a value of $\lambda$, we can determine all $E_{1n}$. Subsequently, we can solve for $x_n$ as a linear programming problem, using Eqs.(\ref{eq:solveLag1})-(\ref{eq:solveLag3}). By exploring the maximum of $\langle\sigma v\rangle$ for all $\lambda$ within the range $[\min(g),\max(g)]$, we can establish the upper limit of fusion reactivity, along with the corresponding values of $x_n$ and $E_{1n}$. Further justification for the assumption of $f_1(v)$ comprising several beams is provided in \ref{sec:f1not}.

\section{Results}\label{sec:result}

{ 
We can begin by gaining some analytical insights. It is evident from Eq. (\ref{eq:solveLag3}) that for any $E_1$, we must have $\langle\sigma v\rangle_{\rm max} \leq \sum_{n=1}^{N}x_n\langle\sigma v\rangle_{Gn,\max} = \langle\sigma v\rangle_{Gn,\max}$.

For $E_1 > E_{1c}$, where $E_{1c}$ represents energy with the maximum $\langle\sigma v\rangle_{Gn}$, we can set $x_1 \to 1$ and $x_2 \to 0$ with $E_{11} = E_{1c}$ and $E_{12} \to \infty$ to satisfy Eq. (\ref{eq:solveLag2}), i.e., $x_1E_{11} + x_2E_{12} = E_1$. Consequently, $\langle\sigma v\rangle = x_1\langle\sigma v\rangle_{G1} + x_2\langle\sigma v\rangle_{G2} = \langle\sigma v\rangle_{Gn,\max}$, indicating that the maximum reactivity approaches a constant, i.e., $\langle\sigma v\rangle_{\rm max} = \langle\sigma v\rangle_{Gn,\max}$ for $E_1 \geq E_{1c}$.

We now numerically calculate the theoretical upper limit { versus $E_1$ for different $T_2$} using the method outlined in subsection \ref{sec:Lag}. It is crucial to handle the points where $E_{1n}=0$ and $E_{1n}\to\infty$ with care. In the numerical aspect, the fusion cross-section data $\sigma(E)$ is confined to the range $E\in[E_{\rm min}, E_{\rm max}]$, typically with $E_{\rm min}=1$ keV and $E_{\rm max}=1-10$ MeV. The cross-section is fitted using several segmentation functions\cite{Bosch1992,Tentori2023}. The $\langle\sigma v\rangle$ for all $\lambda$ within the range $[\min(g),\max(g)]$ is discretized, say, into 100 points, and further solved as a linear programming problem. The $\langle\sigma v\rangle_{Gn}$ data is calculated for $E_{1n}\leq E_{1n,\rm{cut}}$ with $E_{1n,\rm{cut}}=1$ MeV for D-T and $E_{1n,\rm{cut}}=4$ MeV for p-B. Hence, we introduce an artificial point $E_{1n}=100$ MeV (much greater than $E_{1n,\rm{cut}}$) to replace the $E_{1n}\to\infty$ condition, where $\langle\sigma v\rangle_{Gn}=0$, in the linear programming problem to avoid numerical inaccuracies at high energy ranges. The $E_{1n}=0$ point is also included in the calculation.}

Figures \ref{fig:solve_maxsgmvlinprog_icase_1_T2_0}, \ref{fig:solve_maxsgmvlinprog_icase_1_T2_10}, \ref{fig:solve_maxsgmvlinprog_icase_4_T2_0}, and \ref{fig:solve_maxsgmvlinprog_icase_4_T2_50} display the theoretical upper limit of $f_{\langle\sigma v\rangle}$ for D-T with $T_2=0$ and $10$ keV and for p-B with $T_2=0$ and $50$ keV. Additionally, the Monte-Carlo results obtained using the method outlined in subsection \ref{sec:mc} are presented. We observe that the Monte-Carlo results are consistently lower than the upper limit, but they closely approach the upper limit within the fusion energy-relevant energy range, specifically $E_1\simeq 10-30$ keV for D-T and $E_1\simeq 100-500$ keV for p-B. In this range, the typical maximum $f_{\langle\sigma v\rangle}$ is around $1.5-5$ for D-T fusion and approximately $2-4$ for p-B fusion. For smaller energy ranges of $E_1$, $f_{\langle\sigma v\rangle}$ can be much larger, exceeding $5$. However, the total fusion reactivity remains relatively insignificant, which is less relevant for fusion energy applications.

\begin{figure}
\centering
\includegraphics[width=8.5cm]{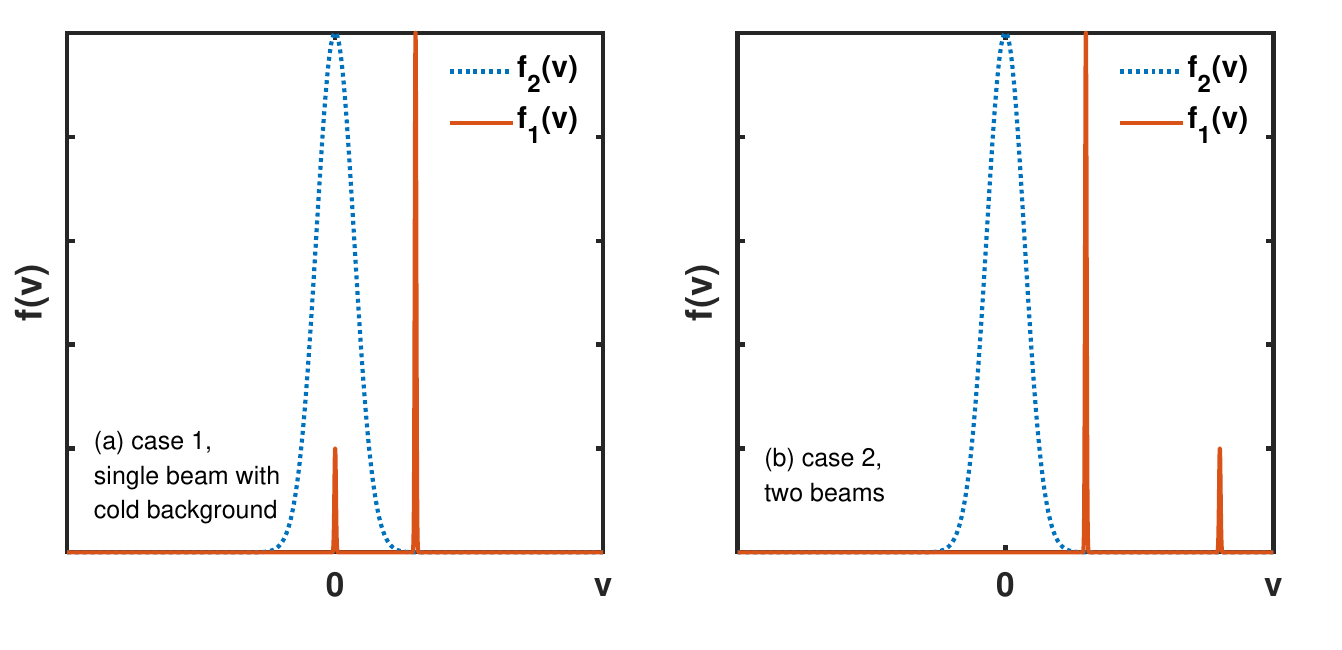}
\caption{Typical distribution functions of $f_1$, which yield maximum fusion reactivity. { For case 2, the second beam can have energy $E\to\infty$ and weight $x\to0$.}}\label{fig:plt_fv}
\end{figure}

\begin{figure*}
\centering
\includegraphics[width=15cm]{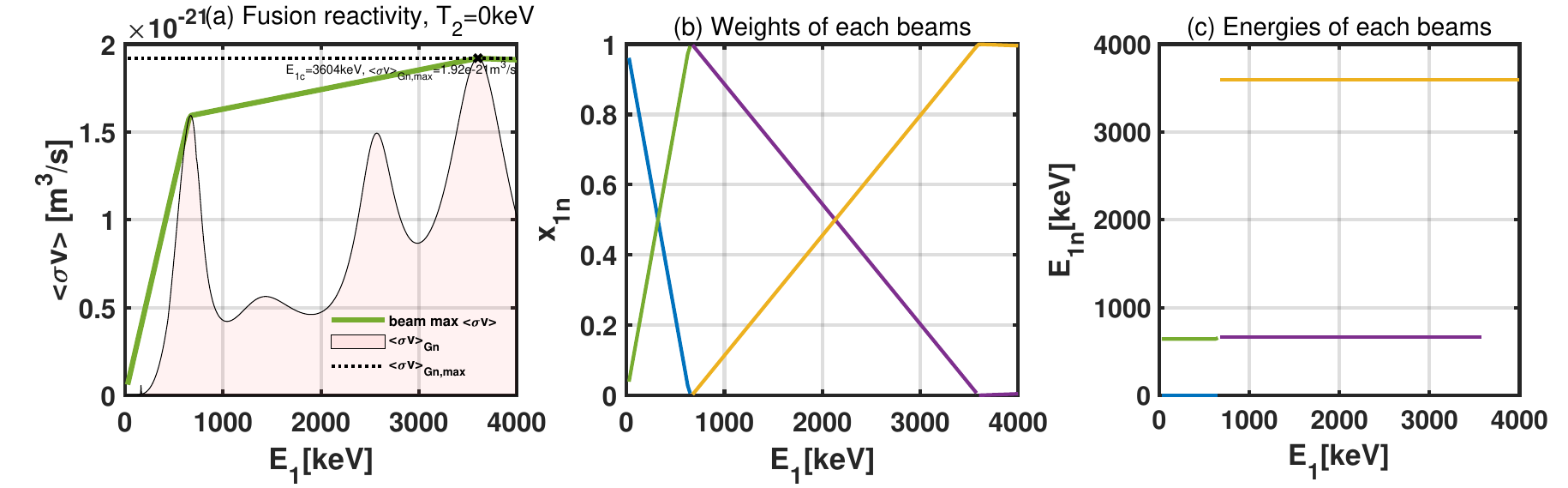}
\caption{{ Upper limit of p-B fusion reactivity ${\langle\sigma v\rangle}$ versus drift energy $E_{1}$, with temperature $T_2=0$, numerically utilizing the method outlined in subsection \ref{sec:Lag}. (a) Fusion reactivity. (b) Weights $x_{1n}$ of each beam for the upper limit case. (c) Energies $E_{1n}$ of each beam for the upper limit case. The $\langle\sigma v\rangle_{Gn}$ versus $E_{1n}$ data is also shown. The calculation is the same as in Fig. \ref{fig:solve_maxsgmvlinprog_icase_4_T2_0}, but with $E_1 \in [0,4000]$ keV instead of $[0,900]$ keV. }}\label{fig:solve_maxsgmvlinprog_icase_4_T2_0_maxE1_4000}
\end{figure*}

{ It is observed that the optimal solution always consists of two beams in Figures \ref{fig:solve_maxsgmvlinprog_icase_1_T2_0}, \ref{fig:solve_maxsgmvlinprog_icase_1_T2_10}, \ref{fig:solve_maxsgmvlinprog_icase_4_T2_0}, and \ref{fig:solve_maxsgmvlinprog_icase_4_T2_50}. The primary beam is chosen to maximize reactivity, while the other beam (cold background or high-energy beam) is selected to satisfy the total energy constraint.} Within the fusion energy-relevant energy range, the results indicate that the maximum $f_{\langle\sigma v\rangle}$ is achieved with a single beam and a cold background (case 1 in Fig.\ref{fig:plt_fv}). For D-T fusion, the optimal beam energy $E_1$ is around 120 keV {(see Figs. \ref{fig:solve_maxsgmvlinprog_icase_1_T2_0}d and \ref{fig:solve_maxsgmvlinprog_icase_1_T2_10}d)}, while for p-B fusion, it is approximately 650 keV {(see Figs. \ref{fig:solve_maxsgmvlinprog_icase_4_T2_0}d and \ref{fig:solve_maxsgmvlinprog_icase_4_T2_50}d)}. These values are close to the energies corresponding to the peaks of { ${\langle\sigma v\rangle}_{Gn}$}. We provide typical distribution functions of $f_1$ that yield the maximum fusion reactivity in Fig.\ref{fig:plt_fv}. To enhance clarity, we also depict the negative of $v$ to visualize the Maxwellian $f_2$. In higher energy ranges, such as $E_1>120$ keV for D-T and $E_1>650$ keV for p-B, two beams yield the maximum fusion reactivity (case 2 in Fig.\ref{fig:plt_fv}). However, this scenario is less pertinent for fusion energy research due to the excessively high energy involved. 

{ We also plot $\langle\sigma v\rangle_{Gn,\max}$ and the corresponding $E_{1c}$ in Figures \ref{fig:solve_maxsgmvlinprog_icase_1_T2_0}, \ref{fig:solve_maxsgmvlinprog_icase_1_T2_10}, \ref{fig:solve_maxsgmvlinprog_icase_4_T2_0}, and \ref{fig:solve_maxsgmvlinprog_icase_4_T2_50} for reference. We observe that indeed $\langle\sigma v\rangle_{\rm max} \leq \langle\sigma v\rangle_{Gn,\max}$ for all $E_1$, and $\langle\sigma v\rangle_{\rm max} = \langle\sigma v\rangle_{Gn,\max}$ occurs for $E_1 \geq E_{1c}$. A difference between D-T and p-B is that there is only one peak in $\langle\sigma v\rangle_{Gn}$ for D-T, but there are multiple peaks for p-B. To showcase the p-B result in an even higher energy region at $E_1=4$ MeV, we plot Fig. \ref{fig:solve_maxsgmvlinprog_icase_4_T2_0_maxE1_4000}. We observe that the optimized beam energies are around the first peak at 650 keV, and the maximum peak $E_{1c} \simeq 3600$ keV. After $E_1 > E_{1c}$, $\langle\sigma v\rangle_{\rm max} = \langle\sigma v\rangle_{Gn,\max}$, consistent with our analytical expectations.}

Another noteworthy observation is that smaller $T_2$ values typically result in higher ${\langle\sigma v\rangle}$ and larger $f_{\langle\sigma v\rangle}$ (cf., compare Fig.\ref{fig:solve_maxsgmvlinprog_icase_1_T2_0}a and Fig.\ref{fig:solve_maxsgmvlinprog_icase_1_T2_10}a). This suggests that a thermal $f_2$ is generally less advantageous for beam-Maxwellian fusion compared to Maxwellian-Maxwellian fusion. This is understandable since the beam-target case can maximize the advantage of the peak cross section. Additionally, it implies that the maximum fusion reactivity occurs when both reactants, 1 and 2, are non-thermal, such as two reactants consisting of beams with opposite drift directions colliding with each other. However, a rigorous proof of this concept falls outside the scope of the present work.

\section{Summary and Conclusion}\label{sec:summ}

In this study, we have determined the maximum fusion reactivity for a non-thermal reactant, specifically deuterium (D) for D-T fusion and proton (p) for p-B11 fusion, in the presence of a thermal Maxwellian background reactant, which is tritium (T) for D-T and boron (B) for p-B11 fusion. Our analysis focused on these two fusion scenarios. We optimized the velocity distribution of the first reactant while keeping its total energy constant, with the goal of achieving the maximum fusion reactivity. The results indicate that the optimized velocity distribution typically comprises one or two beams.

In the context of fusion energy research, we found that the maximum fusion reactivity can often exceed that of the conventional Maxwellian-Maxwellian reactants case by a significant margin, ranging from 50\% to 300\%, within the fusion energy-relevant energy range. The corresponding distribution functions of the first reactant exhibit one or several beam-like features. These findings establish an upper limit for fusion reactivity and provide valuable insights into enhancing fusion reactivity through non-thermal fusion, which holds particular significance in the field of fusion energy research.

It's important to note that this work primarily addresses the mathematical aspects of the problem, and further studies are required to explore the underlying physical implications. For instance, while the cold background of the distribution in Case 1 (Fig.\ref{fig:plt_fv}) may not significantly enhance fusion reactivity, it serves to reduce the average kinetic energy of the system. Additionally, considering the effects of collisions and plasma transport, the actual velocity distributions are subject to change and tend to relax toward thermal distributions. As future research directions, it would be essential to investigate the relaxation of distributions in realistic plasma scenarios in the time scale of confinement time of the plasmas, and explore the implications of having a non-thermal $f_2({\bm v})$.
{ In theoretical terms, while we assert that the results presented in this work represent an upper limit, it's important to note that the demonstration provided here is more akin to a physicist's proof rather than a rigorously formal mathematical proof. A rigorous mathematical proof is beyond the scope of the present work.}


\appendix
\section{Fusion Reactivity for Drift Maxwellian Reactants}\label{sec:sgmvg}
The fusion reactivity of two drift Maxwellian reactants forms the basis for the present work and can be regarded as a type of Green function. Here, we provide details in this appendix. The velocity distribution function is given by:
\begin{equation}\label{eq:fbm}
f_j({\bm v})=\Big(\frac{m_j}{2\pi k_BT_j}\Big)^{3/2}\exp\Big[-\frac{m_j({\bm v}-{\bm v}_{dj})^2}{2k_BT_j}\Big],
\end{equation}
where $j=1, 2$, and $k_B$ represents the Boltzmann constant. The fusion reactivity is given by \cite{Xie2023}:
\begin{eqnarray}\label{eq:sgmvdm}\nonumber
&&\langle\sigma v\rangle_{DM}\\\nonumber
&=&\frac{2}{\sqrt{\pi}v_{tr}v_{d}}\int_0^{\infty}\sigma(v)v^2\exp\Big(-\frac{v^2+v_{d}^2}{v_{tr}^2}\Big)\cdot\sinh\Big(2\frac{vv_{d}}{v_{tr}^2}\Big)dv\\\nonumber
&=&\sqrt{\frac{2}{\pi m_rk_B^2T_rT_d}}\int_0^{\infty}\sigma(E)\sqrt{E}\cdot\\
&&\exp\Big(-\frac{E+E_{d}}{k_BT_r}\Big)\sinh\Big(\frac{2\sqrt{EE_d}}{k_BT_r}\Big)dE,
\end{eqnarray}
Here, $\sinh(x) = (e^x - e^{-x})/2 \simeq x + x^3/6 + \cdots$. The effective temperature $T_{r}$, thermal velocity $v_{tr}$, drift velocity $v_{d}$, and drift energy $E_{d}$ are defined as follows:
\begin{eqnarray}\nonumber
T_{r}=\frac{m_1T_2+m_2T_1}{m_1+m_2},~~v_{tr}=\sqrt{\frac{2k_BT_r}{m_r}},\\
v_{d}=|{\bm v}_{d2}-{\bm v}_{d1}|, ~~E_d\equiv k_BT_d=\frac{m_rv_{d}^2}{2}.
\end{eqnarray}

It's important to note that the drift velocity ${\bm v}_{dj}$ can be in arbitrary directions, meaning ${\bm v}_{d1}$ and ${\bm v}_{d2}$ are not required to be in the same direction.

When ${\bm v}_{dj}=0$, Eq.(\ref{eq:sgmvdm}) reduces to the well-known reactivity for two Maxwellian reactants \cite{Nevins2000}:
\begin{equation}\label{eq:sgmvm}
\langle\sigma v\rangle_M=\sqrt{\frac{8}{\pi m_r}}\frac{1}{(k_BT_{r})^{3/2}}\int_0^{\infty}\sigma(E)E\exp\Big(-\frac{E}{k_BT_{r}}\Big)dE.
\end{equation}

For $T_{1}\to0$, Eq.(\ref{eq:sgmvdm}) reduces to:
\begin{eqnarray}\label{eq:sgmvg}
&&\langle\sigma v\rangle_{G}\\\nonumber
&=&\frac{2}{\sqrt{\pi}v_{t2}v_{d1}}\int_0^{\infty}\sigma(v)v^2\exp\Big(-\frac{v^2+v_{d1}^2}{v_{t2}^2}\Big)\cdot\sinh\Big(2\frac{vv_{d1}}{v_{t2}^2}\Big)dv,
\end{eqnarray}
where $v_{t2}=\sqrt{2k_BT_2/m_2}$, representing the beam-Maxwellian case. The subscript `G' denotes `Green function',  indicating that we use it as the basis for the present work.

If further $v_{t2}\to0$, Eq.(\ref{eq:sgmvg1}) becomes:
\begin{eqnarray}\label{eq:sgmvg1}
\langle\sigma v\rangle_{G}
=\sigma(v_{d1})v_{d1},
\end{eqnarray}
representing the beam-target case.

\section{When $f_1(v)$ is a Continuous Function}\label{sec:f1not}
We demonstrate that when $f_1(v)$ is a continuous function, a solution does not exist. Therefore, we assume that $f_1(v)$ as a series of delta functions is a reasonable choice. For simplification, we will consider the case when $T_2=0$, which implies that $f_2({\bm v}_2)=\delta({\bm v}_2)$. Consequently, we have $\langle\sigma v\rangle=\int d{\bm v}\sigma(v)vf_1({\bm v})$.

We start with the normalization condition:
{\small
\begin{eqnarray}
\int d{\bm v}f_1({\bm v})=\int_0^{\infty}dv\int_0^{\pi}d\theta\int_0^{2\pi}d\phi v^2\sin\theta f_1(v,\theta,\phi)=1.
\end{eqnarray}
}
We redefine:
\begin{eqnarray}
f_1(v)=\int_0^{\pi}d\theta\int_0^{2\pi}d\phi v^2\sin\theta f_1(v,\theta,\phi),
\end{eqnarray}
which yields:
\begin{eqnarray}
\int_0^{\infty}dv f_1(v)=1,\\
E_1=\frac{1}{2}m_1\int_0^{\infty}dv v^2f_1(v)=\frac{1}{2}m_1v_{t1}^2,\\
\langle\sigma v\rangle=\int_0^{\infty}dv\sigma(v)vf_1(v).
\end{eqnarray}

Now, we need to find $f_1(v)$ to maximize $\langle\sigma v\rangle$. Assuming $f_1(v)$ is a continuous function, we can expand it using orthogonal basis functions:
\begin{eqnarray}\label{eq:fhn}
f_1(v)=\sum a_nh_n(v),
\end{eqnarray}
with:
\begin{eqnarray}
\int_0^{\infty}h_n(v)h_m(v)dv=\delta_{mn}.
\end{eqnarray}

Hence:
\begin{eqnarray}
\sum_n a_n\int_0^{\infty}h_n(v)dv=1,\\
\sum_n a_n\int_0^{\infty}dv v^2h_n(v)=v_{t1}^2,\\
\langle\sigma v\rangle=\sum_na_n\int_0^{\infty}dv\sigma(v)vh_n(v).
\end{eqnarray}

We define the Lagrangian:
\begin{eqnarray}\nonumber
&&L(a_1,a_2,\cdots,a_n,\lambda_1,\lambda_2)\\\nonumber
&=&\langle\sigma v\rangle-\lambda_1\Big[\int_0^{\infty}dv f_1(v)-1\Big]-\lambda_2\Big[\int_0^{\infty}dv v^2f_1(v)-v_{t1}^2\Big]\\\nonumber
&=&\sum_na_n\int_0^{\infty}dv\sigma(v)vh_n(v)-\lambda_1\Big[\sum_n a_n\int_0^{\infty}h_n(v)dv-1\Big]\\
&&-\lambda_2\Big[\sum_n a_n\int_0^{\infty}dv v^2h_n(v)-v_{t1}^2\Big].
\end{eqnarray}
This requires:
\begin{eqnarray}
&&0=\frac{\partial L(a_1,a_2,\cdots,a_n,\lambda_1,\lambda_2)}{\partial a_n}\\\nonumber
&=&\int_0^{\infty}dv\sigma(v)vh_n(v)-\lambda_1\int_0^{\infty}h_n(v)dv-\lambda_2\int_0^{\infty}dv v^2h_n(v).
\end{eqnarray}
It turns out that there are no values of $a_n$ that can satisfy the above equation, which implies that the assumption in Eq.(\ref{eq:fhn}) is unreasonable. In other words, $f_1(v)$ should not be a continuous function.


\begin{thebibliography}{99}


\bibitem{Atzeni2004} S. Atzeni and ter-Vehn, Jürgen Meyer, The Physics of Inertial Fusion: Beam Plasma Interaction, Hydrodynamics, Hot Dense Matter ,Oxford University Press, 2004.

\bibitem{Clayton1983} Donald D. Clayton, Principles of Stellar Evolution and Nucleosynthesis, The University of Chicago Press, 1983.

\bibitem{Hartouni2023} E. P. Hartouni, et al, Evidence for suprathermal ion distribution in burning plasmas, Nature Physics, 19, 1, 72-77 (2023).

\bibitem{Nevins1998} W. M. Nevins, A Review of Confinement Requirements for Advanced Fuels, Journal of Fusion Energy, 17, 1, 25 (1998).

\bibitem{Cai2022} J. Q. Cai, H. S. Xie, Y. Li, M. Tuszewski, H. B. Zhou and P. P. Chen, A Study of the Requirements of p-11B Fusion Reactor by Tokamak System, Code, Fusion Science and Technology, 78:2, 149-163 (2022).

\bibitem{Ochs2022} I. E. Ochs, E. J. Kolmes, M. E. Mlodik, T. Rubin and N. J. Fisch, Improving the feasibility of economical proton-boron-11 fusion via alpha channeling with a hybrid fast and thermal proton scheme, Phys. Rev. E, 106, 055215 (2022).

\bibitem{Mehlhorn2022} T. A. Mehlhorn, L. Labun, B. M. Hegelich, D. Margarone, M. F. Gu, D. Batani, E. M. Campbell and S. X. Hu, Path to Increasing p-B11 Reactivity via ps and ns Lasers, Laser and Particle Beams, 2022, 2355629 (2022).

\bibitem{Rostoker1997} N. Rostoker, M. W. Binderbauer and H. J. Monkhorst, Colliding Beam Fusion Reactor, Science, 278, 5342, 1419-1422 (1997).

{\bibitem{Dawson1981} J. M. Dawson, Advanced fusion reactors. Fusion, part B, edited by: Teller, E. (1981) 465.

\bibitem{Liu2023} M. S. Liu, H. S. Xie et al, ENN's Roadmap for Proton-Boron Fusion Based on Spherical Torus, The 29th Fusion Energy Conference, London, UK, Oct. 16-21, 2023.}


\bibitem{Xie2023} H. S. Xie, M. Z. Tan, D. Luo, Z. Li and B. Liu, Fusion reactivities with drift bi-Maxwellian ion velocity distributions, Plasma Phys. Control. Fusion 65, 055019 (2023).

\bibitem{Kong2023} H. Z. Kong, H. S. Xie, B. Liu, M. Z. Tan, D. Luo, Z. Li and J. Z. Sun, Enhancement of fusion reactivity under non-Maxwellian distributions: effects of drift-ring-beam, slowing-down, and kappa super-thermal distributions, { Plasma Phys. Control. Fusion 66, 015009 (2024).}

\bibitem{Kolmes2021} E. J. Kolmes, M. E. Mlodik, and N. J. Fisch, Fusion yield of plasma with velocity-space anisotropy at constant energy, Phys. Plasmas 28, 052107 (2021).

\bibitem{Putvinski2019} S.V. Putvinski, D.D. Ryutov and P.N. Yushmanov, Fusion reactivity of the pB11 plasma revisited, Nucl. Fusion 59, 076018 (2019).

\bibitem{Xie2023a} H. S. Xie, A simple and fast approach for computing the fusion reactivities with arbitrary ion velocity distributions, Computer Physics Communications, 292, 108862 (2023).


\bibitem{Bosch1992} H. S. Bosch and G. M. Hale, Improved formulas for fusion cross-sections and thermal reactivities, Nuclear Fusion, 32, 4, 611 (1992). 

\bibitem{Tentori2023} A. Tentori and F. Belloni, Revisiting p-11B fusion cross section and reactivity, and their analytic approximations, Nuclear Fusion, 63, 8, 086001 (2023).

{ \bibitem{Sikora2016} M. H. Sikora and H. R. Weller, A New Evaluation of the $^{11}B(p,\alpha)\alpha\alpha$ Reaction Rates, Journal of Fusion Energy, 35, 3, 538 (2016).}



\bibitem{Nevins2000} W. M. Nevins and R. Swain, The thermonuclear fusion rate coefficient for p- 11 B reactions, Nuclear Fusion, 40, 4, 865 (2000).

















\end{thebibliography}

\end{document}